\RequirePackage{fix-cm}
\documentclass[twocolumn,11pt]{svjour3}                     % onecolumn (standard format)
\smartqed  % flush right qed marks, e.g. at end of proof
\usepackage{graphicx}

\pdfoutput=1 % if your are submitting a pdflatex (i.e. if you have
\usepackage[T1]{fontenc} % if needed

\usepackage{graphicx} 
\usepackage{lmodern}
\usepackage{url}
\usepackage{epstopdf}
\usepackage{times}
\usepackage{color}
\usepackage{amssymb}

\usepackage{amsmath}
\newcommand{\kms}{kms$^{-1}$}
\newcommand{\vtrue}{$V^{\rm true}$}
\newcommand{\vtrues}{$v^{\rm true}_z$}

\newcommand{\epk}{$\epsilon_{P_k}$}

\newcommand{\vrec}{$\hat V^{\rm rec}$}
\newcommand{\vrecf}{$V^{\rm rec}$}

\newcommand{\pktrue}{$P^{\rm true}_v(k)$}
\newcommand{\pkobs}{$P^{\rm obs}_v(k)$}
\newcommand{\pkdphi}{$P^{\rm \partial\Psi}(k)$}

\newcommand{\pkrec}{$\hat P^{\rm rec}_v(k)$}
\newcommand{\psitrue}{$\Psi^{\rm gal}$}

\newcommand{\pk}{$P_k$}

\newcommand{\dphih}{$\hat V_{\delta}$}%{$\partial_z\hat\Psi$}
\newcommand{\zobs}{$z_{\rm obs}$}
\newcommand{\zcos}{$z_{\rm cos}$}
\newcommand{\vnoise}{$V^{\rm obs}$}
\newcommand{\dphi}{$V_{\delta}$} %{$\partial_z\Psi$}

\newcommand{\chisq}{$\chi^2$}

\newcommand{\sigmadphi}{$\sigma_{\delta_v}$}

\newcommand{\sigmapkv}{$\sigma^v_{P(k)}$}

\newcommand{\deltag}{$\delta_g$}

\newcommand{\ex}{\bf x}
\newcommand{\yy}{\bf y}

\newcommand{\COne}{$0.08 < z \le 0.18$}

\newcommand{\acorr}{$A^{\mbox{\scriptsize{corr}}}$ }

\newcommand{\apj}{Astrophys.~J.}

\newcommand{\apjl}{ApJ Lett.}

\newcommand{\apjs}{ApJ Suppl.}

\newcommand{\prd}{Phys. Rev. D}

\newcommand{\mnras}{MNRAS}

\newcommand{\aap}{A\&A}

\newcommand{\aj}{A J}
\newcommand{\nat}{Nature}

\newcommand{\physrep}{Physics Reports}

\newcommand{\etal}{{\em et al. }}
\definecolor{light-gray}{gray}{0.55}

\font\FermiSmallfont=cmssq8 scaled 1200
\def\LANLppthead#1#2{
\null 
\begin{center}\vskip -1.0truein{\hbox to 7.5truein {
\hfill
\vbox to 1in {\vfill \FermiSmallfont
 Ê Ê Ê Ê Ê Ê Ê\hbox{#1}
 Ê Ê Ê Ê Ê Ê Ê\hbox{#2}
 Ê Ê Ê Ê Ê Ê Ê\vfill}
}}\vskip-0.0truein\end{center}}%FNALppthead

\begin{document}

\title{\boldmath Reconstructing  the velocity field  beyond the local universe%\thanks{Grants or other notes
}


\author{R. Johnston,  D. Bacon,  L.F.A.~Teodoro, R.~C.~Nichol,  M.~S.~Warren,  and C.~Cress}

\institute{R. Johnston \at
              Department of Physics, University of Western Cape, Belville, Cape Town, South Africa, 
%SAAO, P.O. Box 9, Observatory 7935, Cape Town, South Africa, \\
              \email{rwi.johnston@gmail.com}           %  \\
           \and
           D. Bacon\at
           Institute of Cosmology and Gravitation, Burnaby Road, University of Portsmouth, Portsmouth PO1 3FX, UK, 
           \and	 
           L.F.A.~Teodoro \at
           BAER Inst., Space Science and Astrobiology Division, MS: 245-3, NASA Ames Research Center, Moffett Field, CA 94035-1000, USA,
           \and
	  R.~C.~Nichol
	  \at
	  Institute of Cosmology and Gravitation, Burnaby Road, University of Portsmouth, Portsmouth PO1 3FX, UK, 
	  \and
	  M.~S.~Warren
	  \at
	  Theoretical Division, Los Alamos National Laboratory,  Los Alamos, NM 87545 , USA,
	  \and 
	  C.~Cress 
	  \at
	  Centre for High Performance Computing, CSIR, 15 Lower Hope Rd, Cape Town 7700, South Africa  
	   \at
              Department of Physics, University of Western Cape, Belville, Cape Town, South Africa, 
}

\date{Received: date / Accepted: date}

\maketitle

\begin{abstract}
We present a maximum probability approach to reconstructing spatial maps of the  peculiar velocity field at redshifts  $z\sim0.1$, where the velocities have been measured  from distance indicators (DI) such as $D_n-\sigma$ relations or Tully-Fisher. With the large statistical uncertainties associated with DIs, our reconstruction method  aims to recover the underlying true peculiar velocity field by reducing these errors with the use of two physically motivated filtering prior terms. The first constructs an estimate of the velocity field derived from the galaxy over-density {\deltag} and the second makes use of the matter linear density power spectrum {\pk}. Using $N$-body simulations we find, with an SDSS-like sample ($N_{gal}\simeq33$ per deg$^2$ ), an average correlation coefficient value  of $r=0.55\pm{0.02}$ between our reconstructed velocity field and that of the true velocity field from the simulation. However, with a suitably high number density of galaxies from the next generation surveys (e.g. $N_{gal}\simeq140$ per deg$^2$) we can achieve an average $r=0.70\pm{0.02}$ out to moderate redshifts $z\sim0.1$.  This will prove useful for future tests of gravity, as these relatively deep maps are complementary to weak lensing maps at the same redshift. \quad LA-UR 12-24505

%First keyword 
%\and Second keyword \and More}
\end{abstract}

\section{Introduction}\label{intro}
With the advent of all-sky redshift surveys during the 1990s  \cite{Fisher:1995ApJS..100...69F,Saunders:2000,Willick:1996yCat.7198....0W,Willick:1997.ApJS.109.333,Giovanelli:1998ApJ...505L..91G},  the study of peculiar velocities grew into a thriving industry with sophisticated  velocity and density field reconstruction techniques being developed such as:  POTENT \cite{Bertschinger:1989ApJ...336L...5B,Dekel:1999ApJ...522....1D}, VELMOD \cite{Willick:1997.ApJ.486.629,Willick:1998ApJ...507...64W}, Wiener filtering \cite{Fisher:1995MNRAS.272..885F,Zaroubi:1995ApJ...449..446Z} {the  unbiased minimal variance (UMV) method of  \cite{Zaroubi:2002MNRAS.331..901Z} and the inverse Tully-Fisher (ITF)  method \cite{Nusser:1995MNRAS.276.1391N,Costa:1998MNRAS.299..425D}. These approaches allowed powerful  constraints on the cosmological model  (for a comprehensive review see\cite{Strauss:1995PhR...261..271S}).

Central to the  velocity field  reconstruction methodology is  the acquisition of  peculiar velocity data, which is achieved through  distance indicators - empirical  relationships between two or more intrinsic properties of galaxies, which allow us to  estimate a redshift-independent distance $d$ (e.g. the Tully-Fisher relation for spiral galaxies, or $D_n-\sigma$ relations for elliptical galaxies).  In the simplest scenario, we can use this distance to estimate a peculiar velocity $u$ via the well known relation
\begin{equation}
u = cz_{\rm obs} - H_0 d,
\end{equation}
 where $c$ is the speed of light, $H_0$ is the Hubble constant and $z_{\rm obs}$ is the observed redshift, measured spectroscopically. The scatter in distance indicator relations  introduces a  typical statistical error  for a given galaxy of  $\simeq$20\% in distance at scales $\gtrsim$ 50 Mpc (see e.g.\cite{Bernardi:2003AJ....125.1866B,Springob:2007ApJS..172..599S,Magoulas:2010IAUS..262..376M}).  These large uncertainties coupled with  systematics associated   with  distance indicators have thus far limited their range to reconstructing spatially resolved peculiar velocities typically within $z\sim0.015$ (e.g. \cite{Bertschinger:1989ApJ...336L...5B,Nusser:1994ApJ...421L...1N,Fisher:1995MNRAS.272..885F,Willick:1997.ApJ.486.629,Dekel:1999ApJ...522....1D,Zaroubi:2002MNRAS.336.1234Z,Erdodu:2006MNRAS.373...45E}).
 
Recent surveys have provided large catalogues of measured peculiar velocities  (e.g.  the SFI++ galaxy peculiar velocity survey \cite{Masters:2006ApJ...653..861M}, the 6dF galaxy survey  \cite{Jones:2009MNRAS.399..683J} and the 2MASS  selected Flat Galaxy Catalog 2MFGC \cite{Kudrya:2009Ap.....52..335K}).   There has also been ongoing work extending catalogues of inferred distances. These include  Fundamental Plane studies by   \cite{Hyde:2009MNRAS.396.1171H} which  measured $\simeq$50,000 early type galaxies in SDSS; work by    \cite{Magoulas:2012arXiv1206.0385M} who have published 10,000 near-infrared early types in the 6dF Galaxy Survey out to $z < 0.055$;  and \cite{Saulder:2013arXiv1306.0285S} have recently used $\simeq93,000$ from SDSS-DR8 to calibrate the FP out to $z\simeq0.2$.  

However, we are approaching a new era in galaxy survey science where forthcoming  telescopes such as SKA (and their precursors, MeerKAT and ASKAP through HI measurements), Euclid and LSST will measure properties of billions of objects to an unprecedented scale and depth. It is therefore of great interest to explore the possibility of extending velocity field reconstruction to redshifts that compliment other cosmological probes of the gravitational potential.

In recent years there has been a flurry of activity  to develop new probes of  gravity  on cosmological scales  (e.g.\cite{Jain:2010AnPhy.325.1479J}). Several tests have emerged that involve cross-correlating  observables from spectroscopic surveys, galaxy imaging and the CMB. It is useful to examine how one can constrain quantities in the perturbed FLRW metric. For instance, in the Newtonian gauge,
\begin{equation}
ds^2 = -(1-2\Psi)dt^2 + a^2(t)(1+2\Phi)dx^2,
\end{equation}
where  $\Psi$ and $\Phi$ are the Newtonian potentials that describe the temporal and spatial perturbations to the metric respectively, $a(t)$ is the expansion factor of the Universe and $x$ is in comoving coordinates (e.g. \cite{Mukhanov:1992PhR...215..203M,Ma:1995ApJ...455....7M}).\footnote{In regions with negligible anisotropic stress, that is, where the stress-energy tensor is invariant under spatial rotations, or the three principal pressures are identical, the Einstein equations set $\Phi = \Psi$.} Galaxy clustering measurements are sensitive to the $\Psi$ component,  the weak lensing shear is sensitive to the  potentials ($\Psi + \Phi$),   and the CMB (via the Integrated Sachs-Wolf effect)  is dependent on $d$($\Psi + \Phi$)/$d\eta$, where the derivative is taken with respect to the the comoving time $\eta$ (see e.g.\cite{Zhao:2009PhRvL.103x1301Z,Zhao:2010arXiv1003.0001Z,Bean:2010PhRvD..81h3534B,Reyes:2010Natur.464..256R}).  More recently, redshift distortions have been combined with other gravitational probes to further constrain dark energy and gravity models (e.g. \cite{Guzzo2008:Natur.451..541G,Acquaviva:2008PhRvD..78d3514A,Linder:2008APh....29..336L,Song:2009JCAP...10..004S,White:2009MNRAS.397.1348W,Guzik:2010PhRvD..81b3503G}). 

Our main goal in this paper is to explore the possibility of  reconstructing spatially resolved maps of the  peculiar velocity field beyond the local Universe and out to moderate redshifts $(z\sim0.1)$ that will be complementary to other orthogonal cosmological probes  from e.g. weak lensing, as a new probe of gravity.

Given the  large statistical uncertainties on the distances indicators ($\gtrsim 20\%$) at moderate redshift, we will apply a maximum probability approach, making use of  two physically motivated prior terms to weakly regularise and filter the reconstruction. The first uses an estimate of the velocity field derived from the galaxy over-density {\deltag} and the second makes use of the matter density power spectrum $P_\delta$. Through the use of $N$-body simulations we demonstrate that, with measurements with a realistically high signal-to-noise, we can successfully reconstruct the velocity and gravitational potential field out to $z\lesssim0.1$. 

\medskip The structure of this paper is as follows.
  In \S\ref{sec:VMD} we summarise our approach to creating  the input velocity maps.  Details of the velocity reconstruction algorithm are presented in \S\ref{sec:recon} together with results of  applying this reconstruction to our simulations. We provide a summary of these results in \S\ref{sec:results}.  
Finally, in \S\ref{sec:conc} we provide a summary and  discussion.

\section{Simulating the Peculiar Velocity Field}\label{sec:VMD}
 
\noindent Here we describe the procedure used to create simple models of noisy input velocity fields from  mock catalogues, in order to demonstrate the concept of this paper.

\subsection{Peculiar velocities from distance indicators}
 Estimating an accurate  peculiar velocity from current distance indicators (DI) remains a challenging task.  In this section we discuss some of the key sources of uncertainty associated with obtaining a redshift independent distance measurement.
For low redshifts, the radial peculiar velocity component $u(\mathbf{r})$ at the position $\mathbf{r}$ of a galaxy moving with velocity $\mathbf{v}(\mathbf{r})$ is given by 
the relation 
\begin{equation}
u(\mathbf{r}) \equiv \left[\mathbf{v}(\mathbf{r}) - \mathbf{v}(\mathbf{r=0})\right ]\cdot \hat{\mathbf{r}} =
c z -H_0 r\,,
\label{eq:maineq}
\end{equation}
where $c$ is the speed of light, $z$ is the observed redshift for the galaxy, 
$H_0$ is the Hubble constant,  $r \equiv \left| \mathbf{r} \right|$ is the true distance to the object, $\mathbf{v}(\mathbf{r=0})$ denotes the the observer's velocity assumed to be at $\mathbf{r=0}$, and $\hat{\mathbf{r}}$ represents the unit vector along the object's position vector $\mathbf{r}$.  The last equality in the above equation is only valid for  redshifts  $\lesssim 0.3$ (see e.g. \cite{Davis:2001AIPC..555..348D}), which is the regime considered in this paper.  If we are to use the relationship above to estimate $u$, we require not only the measurement of galaxy redshifts but also the inference of a redshift-independent  galaxy distance, the true value of which is $r$. 

The inferred distance is denoted by $d$ and is obtained by means of a DI; empirical  relationships between two or more intrinsic properties of galaxies,  some of which are distance-independent (e.g. velocity dispersion $\sigma_v$, and Luminosity $L$) and some others that are distance dependent (e.g. diameter $D$). The observables used to characterise galaxies differ according to their galaxy type. Thus, distance indicators include the Tully-Fisher relation  \cite{Tully:1977A&A....54..661T} for spiral galaxies, and the Fundamental Plane  (or one of its variations, e.g. the $D_n-\sigma$ relation), for elliptical galaxies
\cite{Dressler:1987ApJ...313...42D,Djorgovski:1987ApJ...313...59D}.

The uncertainties due to the internal scattering in the DI, $\sigma_d$, leads to large uncertainties in the galaxy distances of  $\sigma_d/ d \simeq 0.2$.  One  uses Equation~\ref{eq:maineq}  with  the inferred distance $d$, in order to calculate the peculiar velocity. Note that this requires us to adopt a particular cosmology (i.e. the cosmological parameters assumed must be either stated or marginalised over).  Moreover, the uncertainty in $d$ will propagate via Equation~\ref{eq:maineq} into a large uncertainty on the peculiar velocity for a given galaxy. From Equation~\ref{eq:maineq}  the error on the peculiar velocity is the sum in quadrature of the uncertainties in $cz$ and $H_0d$ since these two variables are measured independently: 
\begin{equation}\label{equ:errorvel}
\sigma_{u} =  \sqrt{\sigma^2_{cz}+\sigma^2_{H_0d}}\,\approx\,\sigma_{H_0d}
\approx H_o \sigma_d = H_0 \times \left(0.2\,d \right),
\end{equation}
assuming that typically $\sigma_{cz}$ will be negligible compared to $\sigma_{H_0d}$.  Hence for instance, a galaxy at $d~=~150$~Mpc\,$h^{-1}$ has $\sigma_u\simeq 3,000$ km\,s$^{-1}$  while its typical peculiar velocity is $u \sim 300$ km\,s$^{-1}$.  It is therefore necessary to average over many galaxies to obtain reasonable signal-to-noise for a mean velocity signal.   However,  peculiar velocity work has long been plagued by  biases stemming from both systematics from distance indicators in the form of Malmquist bias   (e.g. \cite{malmquist:1924,Lynden-Bell:1988,Hendry:1994ApJ...435..515H,Strauss:1995PhR...261..271S})  and calibration bias of  distance indicators. In this work we ask, if systematics are successfully corrected for, would it be possible to then overcome the large statistical errors that would result from studying velocities out to moderate redshifts?  To answer this we explore whether it is possible to recover the velocity field from data with realistic noise levels, assuming that sources of systematics are well understood and accounted for.

\subsection{The observed velocity field}\label{sec:noisevel}
This study is concerned with extracting the peculiar velocity signal from a statistically  noisy measurement that would be measured from e.g. a $D_n-\sigma$ DI at redshifts of order $z\sim0.1$. We are interested in the typical signal-to-noise one needs to successfully reconstruct a peculiar velocity map; we therefore consider two cases where one has either measured a high or low  number density of peculiar velocities. 

 To do this we use a simulated set of galaxies within a CDM cosmological cube that is 384 $h^{-1}$ Mpc on the side \cite{Warren2006ApJ...646..881W}. We define $r$  as  the true (e.g. co-moving) distance  of the object from the observer, and the  line of sight peculiar velocity for an individual object is referred to as the true  velocity {\vtrues}. This set of true velocities represents the ideal peculiar velocity since no statistical noise has yet been added.

To create a catalogue of simulated observed  peculiar velocities (in the absence of selection effects) we firstly compute the distribution of inferred distances $d$ given a set of true distances $r$ within each realisation of the  mock catalogues.  If we have Gaussian errors in the magnitudes obtained from our DI, which are used to compute the inferred distances, then the errors in the inferred distances are lognormal.  The conditional probability density distribution that the inferred distance is between $d$ and $d + {\rm d}d$ given the true  distance $r$, is given by

\begin{equation}\label{equ:cond_dist_true_inferr}
\mbox{Pr} \left( d |r \right)  {\rm d} d=  \frac{{\rm d} \left(\ln d \right)}{\sqrt{2\pi \sigma_d^2} }
\exp \left( - \frac{\left[ \ln r- \ln d  \right]^2 }{2\sigma_d^2}  \right),
\end{equation}
where $\sigma_d$  is a measure of the fractional distance  uncertainty of the Distance Indicator \cite{Landy:1992ApJ...391..494L,Strauss:1995PhR...261..271S}. Throughout our analysis we assume $\sigma_d/d$=0.3 to propagate into our mock observed velocity field.  For a given galaxy in our simulation we can define an observed redshift {\zobs} to be
\begin{equation}
z_{\rm obs} = \frac{H_0\  r +\  v_z^{pec}}{c}
\end{equation}
where, $H_0r/c =$ {\zcos}, the cosmological redshift.

 We now wish to average the data on a particular scale. This is achieved by  pixelising the cube equally in {\ex}, {\yy}  and  redshift {\zobs} with a transverse  pixel  scale of (24 $h^{-1}$  Mpc)$^2$. Thus we have $16\times16\times16$  = 4096 pixels in total. This scale is sufficiently large to smooth out non-linearities  in the velocities. 
We create two galaxy samples which both cover a range  $0.08 < z \le0.18$ but which have different number densities of objects. For the high S/N case we have a total of 300,000 galaxies which equates to a  number density of $N_{gal}\simeq140$ per deg$^2$. In the  lower S/N sample we use  the estimated number density from the SDSS Legacy survey,  as detailed in \cite{Ahn:2012arXiv1207.7137S}, with a total number of  70,000 objects ($N_{gal}\simeq33$ per deg$^2$).  Throughout we will refer to the high and low S/N catalogues as Mock~1 and Mock~2 respectively. For our method we use the observed redshift as a proxy for distance and so bin the data in observed  redshift space. Therefore, for  each slice in \zobs,  we evaluate the average  velocity $V_{ijk}$ for the ($i$, $j$)-th pixel in the $k$-th redshift slice. We denote the  true and observed velocity fields  respectively as {\vtrue} and {\vnoise}. 
 
\section{The Velocity Reconstruction Method}\label{sec:recon}

In this section we describe our velocity reconstruction approach.  The task of a reconstruction algorithm is to infer a good estimate of the underlying true quantity $T$. The probability of a particular hypothesis for the inferred quantity, $\hat{T}$, is to be calculated taking into account the noise of the observations, $N$, and the data $D$. A  common approach is to parameterize  $\hat{T}$  and try to extract such parameters from the observed data, $D$, via Bayes' theorem
\begin{equation}
{\rm Pr}( \hat{T} | D ) \equiv \frac{{\rm Pr} ( D|\hat{T} ) \times {\rm Pr}(\hat{T}) }{{\rm Pr} (D)}.
\end{equation}
${\rm Pr} ( D|\hat{T} )$ is the likelihood while ${\rm Pr}(\hat{T})$ denotes the prior, which contains any previous information or prejudice we have regarding the value of $\hat{T}$. Since $D$ is measured before it is used to infer $\hat{T}$,  ${\rm Pr} (D)$ is a constant. In our algorithm, the relevant data are the noisy measurements of the radial velocity field $V^{\mbox{\scriptsize{obs}}}$  and the quantity we seek to infer is the reconstructed radial velocity field $\hat{V}^{\mbox{\scriptsize{rec.}}}$. Finding the most probable $\hat{V}^{\mbox{\scriptsize{rec}}}$ is an application of maximum probability reconstruction to velocity distance measures; much work has been done in applying this methodology in lensing e.g. \cite{Seitz:1998AA...337..325S,Bridle:2000astro.ph.10387B,Marshall:2002MNRAS.335.1037M,Szepietowski:2013arXiv1306.5324S} and to large-scale structure e.g.  \cite{Kitaura:2009MNRAS.400..183K,Jasche:2010MNRAS.409..355J}.  In our approach the prior will be crucial to reduce noise in the reconstruction. With a particular choice of prior which we will motivate below, the previous equation can be cast as

\begin{eqnarray}\label{equ:lglike}
\nonumber
\log[{\rm Pr}(\hat{V}^{\mbox{\scriptsize{rec.}}}|V^{\mbox{\scriptsize{obs}}})] 
&=&  \log[{\rm Pr}(V^{\mbox{\scriptsize{obs}}}| \hat{V}^{\mbox{\scriptsize{rec.}}})] +
\nonumber \\
&& \log[{\rm Pr}(V|V_\delta)] +\nonumber \\ 
&&  \log[{\rm Pr}(P_V({\rm k)}|P^{\mbox{\scriptsize{th}}}_V(\rm k))]  - \mbox{const.},
\end{eqnarray}
where  the quantity $V_\delta$ is an expected velocity field derived from the galaxy over-density.  $P_V(k)$ is the measured velocity power spectrum and $P_V^{\mbox{\scriptsize{th}}}(k)$ is its theoretical counterpart.  In broad terms, the $V_\delta$ prior term is stating that we (weakly) prefer velocity fields that are in keeping with the velocity field expected from linear theory given a density field. In addition, the $P_V (k)$ prior term acts to prefer velocity fields which have a power spectrum reasonably close to theoretical predictions for the velocity power. Together, these terms state that the prior probability of a reconstructed field is treated as proportional to an assigned probability that this field would deviate from the linear theory prediction, multiplied by the probability that the field's behaviour would deviate from from our preferred power spectrum. Since the permitted deviations in each case are very large, these two terms are treated as independent. The meaning and construction of the prior terms will be  discussed in detail in the following subsections. 
The second and third terms of the previous equation are the logarithm of the prior  while the first one is the logarithm of the likelihood:
\begin{equation}\label{equ:chi2}
 - 2\log[{\rm Pr}(V^{\mbox{\scriptsize{obs}}}| {\hat{V}}^{\mbox{\scriptsize{rec.}}})] \equiv    \chi^2_k=\sum_{i,j=1}^{N_{\mbox{\scriptsize{pix}},k}}\frac{(\hat{V}_{ijk}^{\rm rec}-V_{ijk}^{\rm obs})^2 }{\sigma_{v,ijk}^2},
\end{equation}
where $N_{{\rm pix}, k}$ is the number of pixels in the $k$-th $z_C$ slice. In the likelihood term, we  find  the covariance to be strongly diagonal due to the dominance of noise.  To confirm this, we took one of our simulated galaxy cubes at {\COne} \ and generated 10,000 random realisations. For each realisation we add a new set of random errors to  the distances, with $\sigma_d/d=0.3$. The resulting covariance is shown below in Figure~\ref{fig:cov_vobs}; as expected, the matrix is strongly diagonal.
 We do not at this stage (as one could, as in e.g. the VELMOD approach) include systematic effects and biases into our likelihood analysis; here we are interested in a proof-of-concept approach to probe the {\it high} redshift regime.

\begin{figure}
	\includegraphics[width=0.49\textwidth]{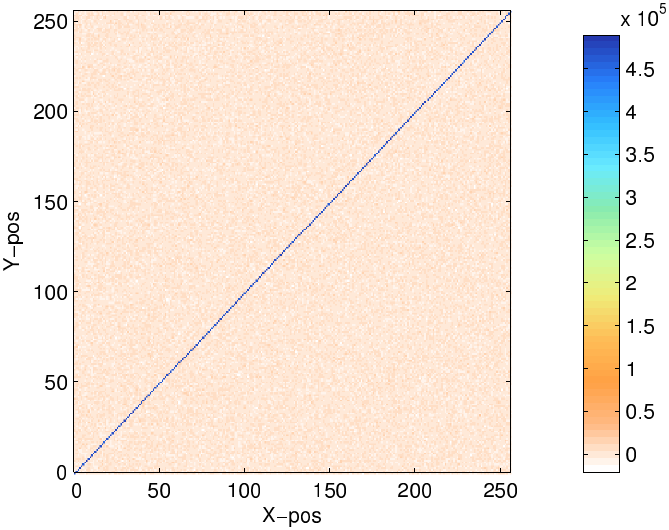}\hfill
        \caption{\small Covariance of the $V_{obs}$ field as calculated from 10,000 realisations of the distance errors on the catalogue galaxies. }
        \label{fig:cov_vobs}
  \end{figure}

\begin{figure*}
%\begin{center}
	\includegraphics[width=0.34\textwidth]{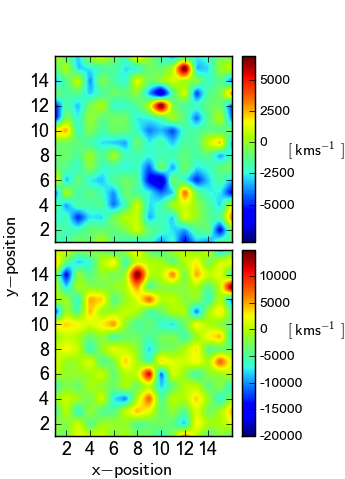}\hfill
	\includegraphics[width=0.62\textwidth]{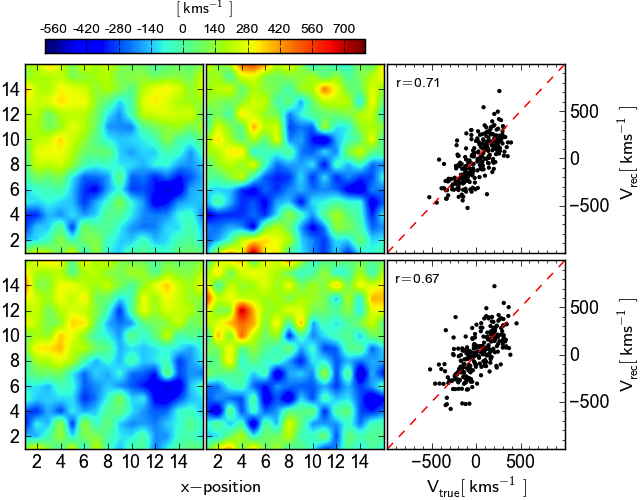}	

        \caption{\small Velocity reconstruction maps. The top row shows results for a mock sample with a  number density of galaxies $N_{gal}\simeq140$ per deg$^2$ while the bottom row is for a  mock sample with a number density of $N_{gal}\simeq33$ per deg$^2$. For this example we show the results for a redshift slice of  $0.118<z<0.124$. 
        For each column going from left to right we show the input noise velocity maps,  the true radial velocity field {\vtrue}, the reconstructed velocity map, and  the resulting correlation between the  two maps r[{\vtrue},{\vrec}] respectively. The dashed line is a one-to-one relation.}
%        \end{center}
        \label{fig:sdssrec}
\end{figure*}

  \begin{figure*}
  	\includegraphics[width=0.48\textwidth]{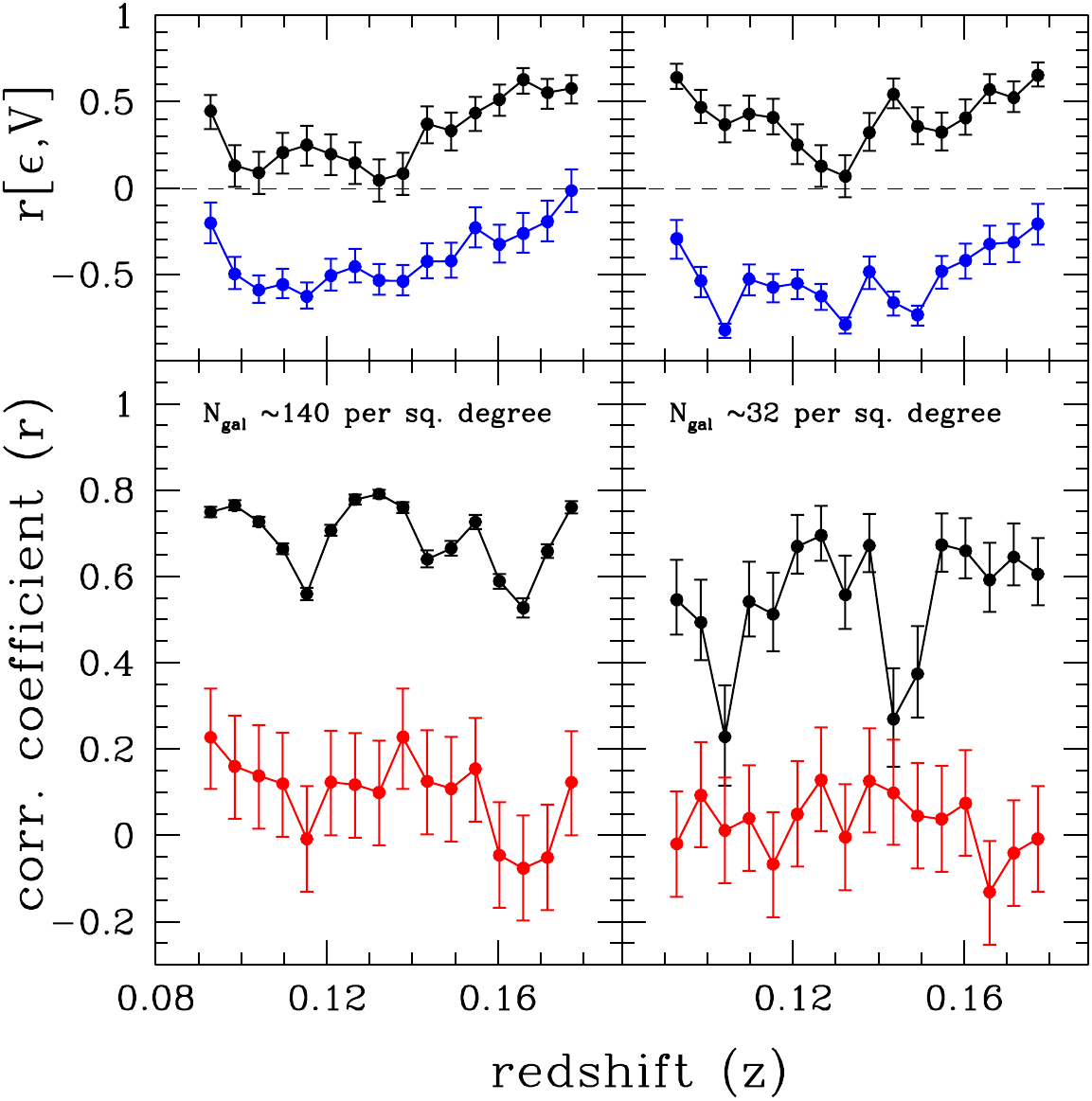}\hfill
	\includegraphics[width=0.48\textwidth]{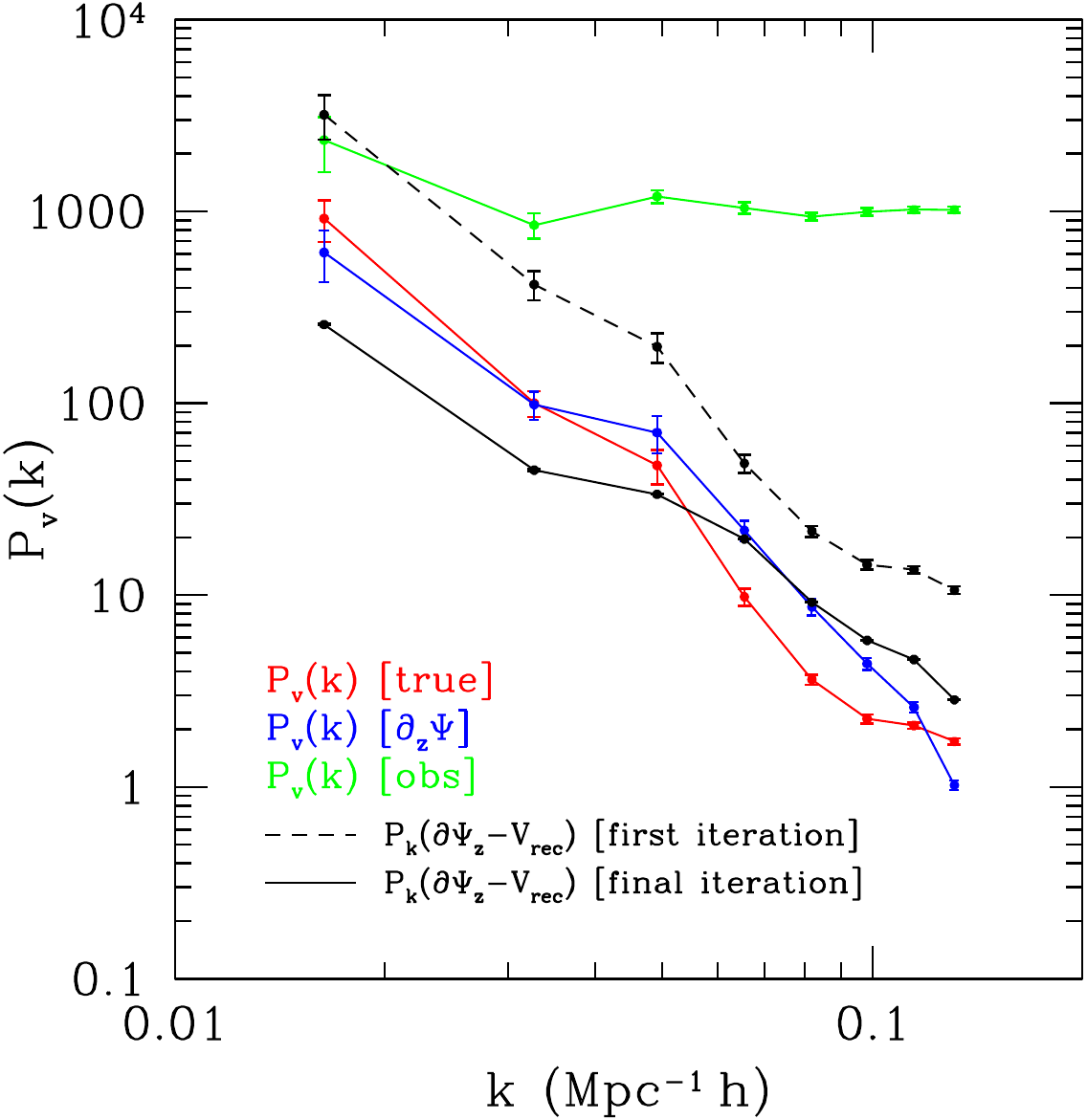}
        \caption{\small {\it Left:} Correlation coefficient values for the final reconstruction of the peculiar velocity field.  The bottom panels show the correlation between r[{\vnoise},{\vtrue}] (red points) for each redshift slice Mock~1 (left) and Mock~2 (right). The black points show the mean correlation value for the 1000 MC samples and the true velocity field i.e. r[{\vrec},{\vtrue}].  The top panels show the respective correlation coefficient between the residuals $\epsilon$ (where $\epsilon$~=~{\vtrue}~-~{\vrec})  and the {\vtrue} (black points) and {\vrec} (blue points). {\it Right:} 
        The error power spectrum compared for the various velocity fields. The red line shows the power spectrum of the true velocity field {\vtrue}, the blue line is the power spectrum on the {\dphi}$^{\rm gal}$ field (computed from the galaxy over density) and the green line is the power spectrum of the observed noisy input velocity map {\vnoise}. The dashed black line shows the residual power of ({\dphi}$^{\rm gal}$ - {\vrec}), where {\vrec} is the reconstructed velocity field for the first $n=1$ iteration of the algorithm. The solid black line is then the residual power for the final iteration of the algorithm at $n=20,000$.  }
        \label{fig:rho}
\end{figure*}

\subsection{Approximate form for likelihood}

We note that in its current form,  Equation~\ref{equ:chi2} is inadequate: the \chisq distribution is correct for data with Gaussian errors; however, as shown in Expression~\ref{equ:cond_dist_true_inferr} our inferred distances $d$ are drawn from a lognormal distribution which is subsequently propagated  into the observed velocity field \vnoise. Since there are many galaxies in each map pixel, the error on the mean velocity in the pixel will be Gaussian, but the lognormal distribution for each galaxy introduces an offset in the mean velocity. We therefore introduce a  correction  to account for this mean offset. Recall that for each object $i$ we have defined an inferred distance $d_i$ and a true distance $r_i$ which are then converted into effective redshifts \zcos$_{,i}$ and \zobs$_{,i}$ as $H_o d_i/c$ and $H_0 r_i/c$, respectively.  We compute \acorr by generating 100 different catalogues of observed distances, and computing $\langle z_{\rm obs}-z_{\rm cos} \rangle |_{z_{\rm obs}}$ and the dispersion on this for each realisation. Then we have as our corrected likelihood term

\begin{equation}\label{equ:chi2_2}
\chi^2_k=\sum_{i,j=1}^{N_{\mbox{\scriptsize{pix}},k}}\frac{(\hat{V}_{ijk}^{\rm rec}-V_{ijk}^{\rm obs} + A_{\mbox{\scriptsize{corr}},k})^2}{\sigma_{v,ijk}^2}.
\end{equation}
The success of this approximation will be demonstrated in our results section \S~\ref{sec:results} below. Having obtained a form for the likelihood, we now turn to the prior terms.

\subsection{The $V_\delta$ Field Prior Term}
In this subsection we discuss the first prior term used in the reconstruction. We estimate the velocity field derived  from the observed over-density $\delta_g$ of the galaxy distribution. The motivation for such a prior term is to express a modest preference for velocity fields that are consistent with linear gravitational infall given a galaxy density field. 

From this starting point we can  construct an estimate of the gravitational potential field $\Psi$ via the Poisson equation (assuming a linear bias on these scales; we currently set $b=1$, but this could be altered for a particular galaxy sample). We include the required modification to take into account the effects of redshift space distortions; in Fourier space the gravitational potential can be determined by
\begin{equation}\label{equ:phi}
\Psi^{\rm gal}({k}) = \frac{3\Omega_mH_0^2\delta_{z}({k})}{2a({\bf k}^2 + \beta{\bf k_z^2})},
\end{equation}
where $\Omega_m$ is the dimensionless matter-density parameter, $H_0$ is the Hubble constant and {\bf k} is the wave-vector.  We have assumed a value of $\beta=0.309$ as constrained in \cite{Tegmark:2006PhRvD..74l3507T}. The scale factor $a$ is computed at the midpoint of each redshift  slice, given by
$a=1/(1+z)$. Throughout we assume $\Omega_m=0.3$ and $H_0$=72 kms$^{-1}$Mpc$^{-1}$.

The estimated velocity field is obtained using $V_{\delta}^{\rm gal}=2f\partial_z \Psi^{\rm gal} / 3\Omega_m a\simeq\partial_z\Psi^{\rm gal}$. In Fourier space this is  computed by
\begin{equation}\label{equ:dphi}
%FT(\partial_z\Psi^{\rm gal})=i\Psi^{\rm gal}(k) {\mathbf{k}}_z
FT(V_{\delta}^{\rm gal})=i\Psi^{\rm gal}(k) {\mathbf{k}}_z
\end{equation}
Here  ${\mathbf{k}}_z$ is the $z$-component of $\mathbf{k}$. Now we can write the $V_\delta$  prior term in the form
\begin{equation}
\chi^2_{V_\delta} = \sum_{i=1}^{N_{\rm pix}}\frac{(\hat V^{\rm rec} -V_{\delta}^{\rm gal})^2}{\sigma_{V_\delta}}
\end{equation}
The quantity $\sigma_{V_\delta}$ is a parameter determining how much this prior term will  dominate in the reconstruction. The true covariance of  \dphi$^{\rm gal}$ will not be diagonal, as even on degree scales the $\delta$ field is correlated between pixels. Hence if we wished to make a joint reconstruction between the velocity and overdensity fields, we should include this non-diagonal covariance. However, this joint reconstruction would strongly weight the information from the less noisy \dphi$^{\rm gal}$  field over the more noisy velocity field, which is not the point of this paper - we are interested in whether we can use the \dphi$^{\rm gal}$  field as a {\it weakly} informative prior for extracting information from the velocity field.  That is, an important question is whether the final reconstruction is more correlated with the {\dphi} field or the true velocity field {\vtrue} and, therefore, whether we are dominated by the prior. It is for this reason that we choose a large value  for \sigmadphi; the value on the diagonal, for a joint reconstruction, would be the error one makes in estimating \dphi$^{\rm gal}$ within a pixel ($\simeq$50 \kms ) whereas we choose \sigmadphi=1000 \kms; we will show that this satisfies the condition that the prior does not dominate the reconstruction.

\subsection{The Peculiar Velocity Power Spectrum Prior Term}
\noindent
The second prior term makes use of a theoretical velocity power spectrum to regularise the {\vrec} field.  Using this information in reconstructions is not a new concept;  \cite{Kaiser1991ASPC...15..111K} demonstrated the usefulness of adopting a power spectrum prior for regularising non-parametric fits in their work on density field reconstruction from peculiar velocities.

The $V_\delta$ prior from the previous section compares the reconstructed and position-derived velocity fields on a pixel by pixel basis. As we shall see, the introduction of the {\pk} prior  term complements this by regulating the  reconstruction on a variety of scales.

In order to use this prior, we need to obtain a suitable theoretical velocity power spectrum; here we use the power spectrum of the true velocity field {\vtrue} in the simulations, but in principle we could use a range of LCDM power spectrum predictions and marginalise our results over these.  We compare the theoretical velocity power spectrum {\pktrue}  with each trial reconstruction power spectrum {\pkrec} so that the $P_k$ prior term takes the form
\begin{equation}\label{equ:pk}
\chi^2_{P_k} = \sum_{i=1}^{k_{\rm tot}} \frac{[\hat P_v(k)^{\rm rec} - P_v(k)^{\rm theory}]^2 }{[\sigma^v_{P(k)}]^2},
\end{equation}
where $k_{\rm tot}$ is the total number of wave numbers and we can choose
\begin{equation}\label{equ:sigpkv}
\sigma^v_{P(k)} = aP_v(k)^{\rm theory}.
\end{equation}
The  parameter  $a$ in Equation~\ref{equ:sigpkv} allows us to modify the strictness of this prior term in the reconstruction. We have found that the off-diagonal terms  are small  and therefore can be neglected in the above prior term.  Note that  Equation~\ref{equ:pk} assumes the power spectrum has Gaussian errors; this is valid except on the largest scales.

\begin{figure*}
   	\includegraphics[width=1.\textwidth]{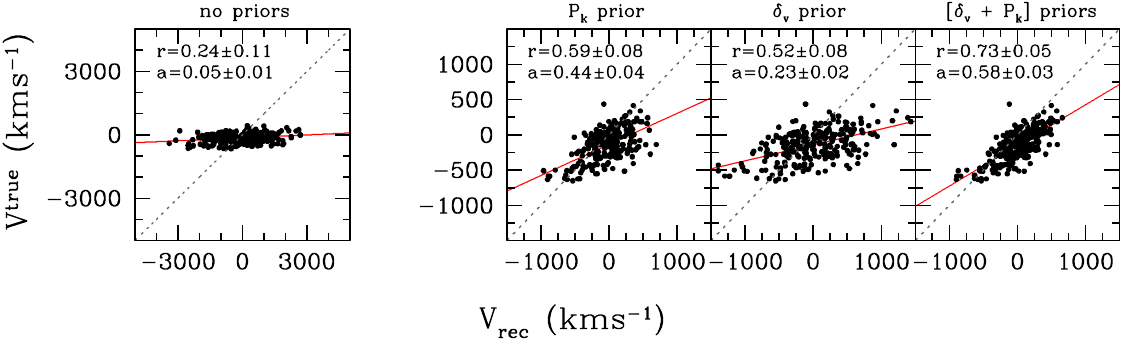}
        \caption{\small An example of the effect of the different prior terms on the reconstruction at a given redshift slice $ 0.101<z\le0.107$. The left  panel shows the final reconstruction of the velocity field where no prior terms have been activated. The second panel  shows the final reconstruction when only the $P_k$ power spectrum prior term is applied with  {\sigmapkv}=0.1{\pktrue}. The third shows the reconstruction with the  baryon $V_\delta$ prior term only with $\sigma_{V_\delta}$=1000.0 {\kms}.  Finally, the fourth shows the combination of the two prior terms. }\label{fig:priors_rec}
\end{figure*}

\subsection{Implementation}
In this section we briefly outline how we implement our algorithm. 
\begin{enumerate}
\item The likelihood is maximised iteratively by beginning with $N_t$ initial trial velocity fields (we choose $N_t=15$).  For each iteration of the reconstruction  we allow a random walk for   $n=2,N_t$  trial reconstructions by adding an amount of uniform random noise  with  of magnitude given by  $\sqrt{P^{\rm theory}(k)}$ to each mode. The $n=1$ field is not perturbed [see step (ii)].
\item We compute the log probabilities for each $n$th trial reconstruction. We determine which of the trials has the minimum  log probability.  If the trial with minimum log probability corresponds to  $n=1$, we resample from the parent sample in the next iteration of the code and allow the pixel variation  to reduce for the new set of trial fields. However, if  the minimum log probability corresponds to $n \ne 1$, we set the corresponding trial {\vrec} to be the new template and feed this back in to  Step (i), where a new set of trial velocity fields are generated based on this new template. This process is repeated until convergence is reached.
\end{enumerate}

\section{Results}  \label{sec:results}
We have applied our reconstruction technique to two simulated galaxy cubes (Mock 1 and Mock 2)  over a  redshift range $0.09<z\le 0.18$; the galaxy cubes have  average number density of galaxies of  $N_{gal}\simeq140$ and $\simeq33$ per deg$^2$ $N_{gal}$ respectively.  Figure~\ref{fig:sdssrec} shows velocity maps  for a selected redshift slice at $0.118 < z\leq 0.124$.  The top row represents the results for Mock 1, while the bottom row shows results for Mock 2.  From left to right we show the input observed noisy velocity map {\vnoise}, the true velocity field {\vtrue}, the final reconstructed velocity field  {\vrec}, and finally the correlation between {\vtrue} and {\vrec}.  

In both mocks we can see that, due to the observational uncertainty $\sigma_d/d=0.3$ we propagate into the observed velocity field {\vnoise}, the observations are completely noise dominated. Comparing both maps to the input true velocity field we find little correlation between them. By applying our method including the two  terms contributing to our prior, we can see that the final reconstructed fields correlate rather well with the {\vtrue}. Mock~1 ($N_{gal}\simeq140$ per deg$^2$)  shows an overall  better reconstruction ($r=0.71\pm0.05$)  compared to Mock~2 which has considerably lower number density of objects ($r=0.67\pm0.08$). However, in both cases with this example we can see that the main large scale structures are recovered well.  

In Figure~\ref{fig:rho}, left panel  we show the correlation between {\vrec} and {\vtrue} for each redshift slice in the mocks. In the lower part of this panel we show correlation coefficient values $r$ between [{\vtrue},{\vnoise}] (red points) and [{\vtrue},{\vrec}] (black points) for Mock~1 (bottom  left) and\ Mock~2 (bottom right). In the upper part of the panel we show the corresponding correlation  coefficient values between the residuals $\epsilon$  and {\vtrue} (black points) and {\vrec} (blue points).  We observe that the overall correlation  with the input observed velocity map {\vrec}  across the mocks is predictably low,  with a  maximum $r=0.23\pm0.11$ and a minimum $r=-0.008\pm 0.122$. This  is due to the addition of our large $\sigma_d/d=0.3$ error on the distances at these relatively high redshifts, which can be seen visually in the noise map from   Figure~\ref{fig:sdssrec}.

In contrast, the black points on the lower part of the left panel of  Figure~\ref{fig:rho}  demonstrate that we achieve a good reconstruction within each redshift slice. By comparing  to the underlying true velocity map, we find a correlation of $r=0.69\pm{0.01}$ in Mock~1 (the high S/N case).  We can see that the result for this lower number density of objects (Mock~2) in the right hand upper panel follows a similar  trend to the Mock~1 case. However, as expected there is a  general systematic shift toward lower correlation values, with an average value of $r=0.55\pm{0.02}$ .  In particular, we note an excessive dip in correlation in the Mock~2 with values of $r=0.23\pm0.12$ and $r=0.27\pm0.12$  $z\simeq0.104$ and $z\simeq0.143$ respectively. On closer inspection we find that this  is due to  a few  outlying pixels with extremely high velocities of the order $\sim$6,000~{\kms} which the algorithm has failed to constrain effectively.  Finally we note that in both mocks the correlations between the residuals $\epsilon$ with {\vtrue} and {\vrec} are not consistent with zero, but show a systematic offset; this will need to be improved upon in future studies. Nevertheless, this is an encouraging result demonstrating that it may be possible to apply our technique to existing data from e.g. SDSS. At the very least, forthcoming surveys should provide the required signal-to-noise level to reconstruct the velocity field  at moderate redshifts.

\subsection{The residual power spectrum}\label{sec:priorerr}
 \begin{figure*}
\begin{center}
	\includegraphics[width=0.5\textwidth]{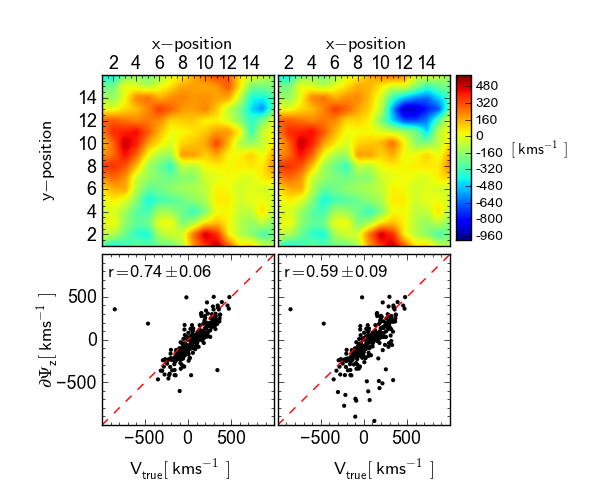}\hfill
	\includegraphics[width=0.5\textwidth]{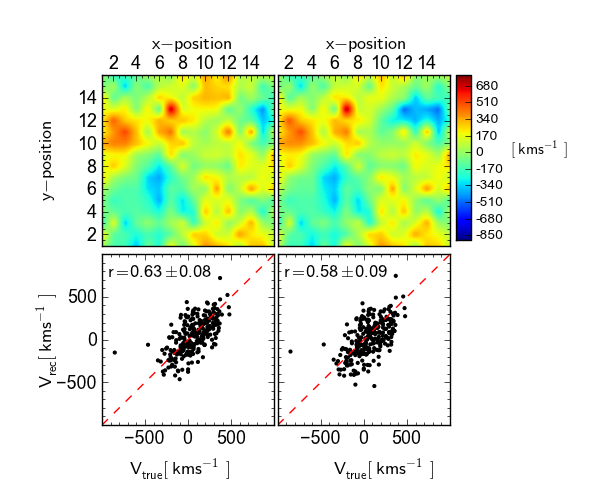}	
        \caption{\small  The left hand panel set shows: the {\dphi} field at $0.135 < z \le 0.141$ (top left) and the same field with an artificial mass embedded; this  can be seen as  the dark blue region in the velocity field. The bottom left panel shows the [\vtrue,\dphi] distribution, where \dphi \ relates to the map directly above. The bottom right panel is the  [\vtrue,\dphi] distribution where  \dphi \ relates to the map with the overdensity embedded in the field. The right hand panel set  shows the reconstructed velocity fields (top row) for the same redshift range - the left hand plots show the case where no artificial mass has been added to the $V_\delta$ prior and the right shows the result where it was been added.  }  \label{fig:blob}
        \end{center}            
\end{figure*}

A useful statistic with which to evaluate the reconstruction is the power spectrum itself.  On the right hand panel of Figure~\ref{fig:rho}  we have plotted the power spectrum as measured from the observed velocity field {\vtrue} shown in green, the true velocity field shown in red, the {\dphi}$^{\rm gal}$ field shown in blue, and finally the residual power $\epsilon_{P_k} ~=~P_k$({\dphi$^{\rm gal}$ - {\vrec}) after  the first iteration of the code (dashed black line) and the final iteration.  

The difference between {\pkobs} and  {\pktrue} is substantial, with the observed power being flat across most scales due to the dominant measur ement noise.  The power spectrum of the true velocity field, {\pktrue} and that of the \dphi\ field {\pkdphi} are rather similar, showing large power on large scales.  The residual power {\epk} reveals important information on how the reconstruction evolves; if we look at {\epk} after the first iteration of the reconstruction then we can see that already the shape of its power spectrum is similar to that of {\pktrue} and {\pkdphi} but with a comparative excess on all scales.  By the final iteration of the code $n=20,000$, {\epk} (black solid line) now has similar power to   {\pkdphi}  on small scales $k\gtrsim 0.06$, but less on scales larger than this.  At $k\lesssim  0.05$ we observe a similar trend when compared to {\pktrue}.  This is a promising result as it shows that the residual power is smaller than the true fluctuations on these scales.
%In the next section we look at the  contribution of the individual  prior terms  to the reconstruction.

\subsection{Exploring the impact of the priors}\label{sec:prioreffect}
The results presented in the previous section use the power spectrum $P(k)$ prior term and the $V_\delta$ prior term in combination. In this section we examine how each prior is affecting the reconstruction process individually.  In Figure~\ref{fig:priors_rec} we show four scenarios of possible reconstruction processes within $ 0.101<z\le0.107$. In all cases we show the [{\vrecf},{\vtrue}] distributions with their respective correlation coefficient values $r$  and the slope of their respective slopes $a$. Firstly, we test the reconstruction method where no priors are applied as shown in the first  panel (left hand side) of the figure.  This serves  as a null test, where we are essentially fitting to the {\it observed} velocity data which is a convolution of the true velocity information and the lognormal noise derived from the simulated observed distances.  As expected, the plot shows a very noisy reconstruction with a correlation of $r=0.24\pm0.11$ and a very broad distribution with velocities in excess of $\sim$3000 {\kms} at the tails. 

In the second panel we show the reconstruction where only the $P_k$ prior has been applied (with the same value for {\sigmapkv} as used in the previous section).  What is immediately clear is a much improved tighter relationship between {\vtrue} and {\vrec} with a correlation of $r=0.59\pm0.08$. We also find an improved slope $a$ of the correlation, $a=0.44\pm0.04$.

The third panel of Figure~\ref{fig:priors_rec} shows the effect of the $V_\delta$ prior which compares the {\vrec} field with {\dphi} field derived from the galaxy over density.  We  adopt the same {\sigmadphi}=1000 {\kms}  value as in our previous analysis. The motivation for choosing such a conservative error on the prior is to limit its effect on regularising the reconstruction and avoid us simply fitting  the {\dphi} field.  As we can see,   the final reconstruction with the $V_\delta$ prior alone  shows an improved but still modest  correlation of $r=0.52\pm0.08$ and a slope of $a=0.23\pm0.02$.

Finally, by combining the two priors as shown in the fourth panel, we find a much improved reconstruction with $r~=~0.73\pm0.05$ and  $a=0.58\pm0.03$, apparently indicating the usefulness of including both priors.  We can compare this correlation  with the  correlation between the reconstruction and the $V_\delta$ prior within the same redshift slice, which is $r=0.63\pm0.08$; the reconstruction is closer to the truth than it is to the prior, which is an important criterion for the success of our method. To understand further whether the reconstruction is a faithful one, it is important to consider the effect of observational biases.
\subsection{Exploring observational bias in the priors}\label{sec:obseff}
Applying our method to real survey data will naturally be subject to various selection effects and biases. Understanding how this might impact the $V_\delta$ prior, and consequently  our velocity reconstruction, is necessary if the method is to be successfully applied to future data-sets.  We have performed further tests of how much the  $V_\delta$ prior will bias  the reconstruction if in some region mass does not follow galaxy counts, i.e. if there is a baryonic feature present which is not proportionally represented in the underlying gravitational potential field. We test this by artificially embedding into the gravitational potential field a `fake' over-dense region which has no counterpart in the true velocity field.  We calculate the resulting {\dphi} field and hence the $V_\delta$ prior term.   The purpose of this test is to observe if such effects will substantially contaminate the final reconstruction of the velocity field. 

In Figure~\ref{fig:blob} we consider the scenario where a large mass would be inferred to be present from the galaxy clustering, modelled by a gaussian embedded into the {\psitrue} field over several redshift slices.  In the left hand panel set of the  figure, we show the  redshift slice where this false mass is most prominent at $0.135 < z \le 0.141$ and is then propagated into the \dphi \ field (top right).  To distinguish between the two velocity fields  we will refer to the  altered prior information with a ``hat" above the observable e.g. {\dphih}. The top  map row in this panel set shows the two \dphi \ maps; we  observe a negative excess in the {\dphih}   velocity potential extending  to $\sim1000$~{\kms}  centred at  [13,13] on the [$X,Y$] pixel plane (top right).   The difference between {\dphi} and {\dphih} is perhaps more clearly shown in the bottom rows where we plot the distributions of both [{\dphi},{\vtrue}] (left) and [{\dphih},{\vtrue}] (right).  In the right-hand panel of this row we now see a skewed distribution of velocity pixels toward the bottom left part of the plot, compared to the {\dphi} on the left where it remains a relatively tight distribution. Comparing  their relative correlations we can see that with the added mass we observe a reduced correlation coefficient from $r$[\vtrue,\dphi]=$0.74\pm0.06$ to  an $r$[\vtrue,\dphih]=$0.59\pm0.09$ corresponding to a $\Delta r= 0.15$. 

Our new {\dphih} field provides a modified $V_\delta$ prior term of the form
\begin{equation}
 \frac{(\hat V_{rec} - \hat V_\delta)^2}{\sigma_{V_\delta}^2}.
\end{equation}
We then carry out our reconstruction method, afflicted with this poor prior.  Comparing the two distributions in the right hand panel set of  Figure~\ref{fig:blob}, we can clearly see that there is very little residual bias in the final reconstruction of the velocity field despite the prominent artificial velocities present in the {\dphih} field.  By comparing the respective correlations we now find for the case where no mass has been added $r$[\vtrue,\vrec]=0.63$\pm$0.08 and the case where the mass has been added to the \dphih field  $r$[\vtrue,\vrec]=0.58$\pm$0.09 corresponding a $\Delta r= 0.05$. This further demonstrates how the $V_\delta$ prior term does not  dominate the reconstruction;  we are extracting useful corrective information from the observed velocity field.

\section{Discussion \& Considerations for the Future}
\label{sec:conc}
In this study we have carried out a first step towards a reconstruction the velocity field  on cosmological scales, through the use of measured peculiar velocities as derived from current distance indicators.  Previous methods have focused on velocity reconstruction within the local Universe. Our motivation is to look toward the possibility of pushing to cosmological depths where such maps of the velocity and gravity field could be used as a cosmological probe of GR. 

To begin developing an approach that can reconstruct resolved maps of the velocity field out to $z\lesssim0.1$, we created two simulated mock catalogues. Both catalogues were over the range $0.09<z<0.1$ with a  scale of  384 $h^{-1}$ Mpc on the side. However, the two mocks had different number densities of objects; the first mock had  a total of  300, 000 objects equating to a number density of   $N_{gal}\simeq140$ per deg$^2$, while the second contained 70, 000 objects ($N_{gal}\simeq33$ per deg$^2$).   The mocks were pixelised  to create a smoothed true velocity field {\vtrue} and a velocity field that represents the noisy observed data, {\vrec}, with a transverse  pixel  scale of (24 $h^{-1}$  Mpc)$^2$. We have modelled a volume-limited sample with a roughly uniform redshift distribution within the redshift range of interest, allowing us to bin the data in equal size redshift bins, in redshift \zobs  space along the line of sight.   In reality one would prefer to maximise the amount of peculiar velocity data available, and a volume limited selection is certainly not optimal.

In this work we have focused only on the contribution of the statistical noise from distance indicator measurements, which then propagate into the observed velocity field. We chose a conservative estimate for the statistical  error on the distances of $\sigma_d/d = 0.3$, which is greater than current error estimates from distance indicators.  However, we do note that  a future development of our approach should include other observational effects/biases into our likelihood analysis in a similar way to that by Willick and Strauss VELMOD approach. Moreover, it will be useful to examine specifically the compounding effects due to DI calibration bias as described in detail in \cite{Strauss:1995PhR...261..271S} by including this in our mocks as well as applying our approach to existing observed data.

Key to our method is the  inclusion in our prior of two terms which guide the velocity reconstruction. The first prior term, featuring {\pk},  regulates the reconstruction on various scales by comparing  a theoretical velocity power spectrum {\pktrue} to the one computed for the trial reconstruction {\pkrec}.  The second prior term, featuring $V_\delta$,  compares the velocity field {\dphi} derived from the galaxy over-density {\deltag} with the trial reconstructed velocity field {\vrec}.  

The  results show that overall we can recover the velocity field to a high degree of correlation with {\vtrue}. For the cube at the  high number density mock,  we found that we could achieve a consistently high correlation coefficient value of $r$[{\vtrue},{\vrec}]~$\simeq0.7$ for  a given slice in redshift. For the mock with a number density closer to that of an SDSS type survey we found we could still reconstruct the main structures but with a lower overall correlation of $r=0.55\pm{0.02}$.  

As a next step, it will be of interest to modify our approach to use the peculiar velocity field to reconstruct the gravitational potential itself at intermediate redshifts of $z\sim0.1$, which will complement gravity maps from other probes.

\section*{Acknowledgements}
We would like to extend special thanks to Enzo Branchini and Adi Nusser for stimulating discussions, their comments  and reading several drafts of this paper. 
We also would like to thank Martin Hendry, Mat Smith, Andreas Faltenbacher, Roy Maartens, Daniele Bertacca, Rafal Szepietowski, Yong-Seon Song, Kazuya Koyama, Prina Patel, Emma Beynon, Robert Crittenden and Philip Marshall for useful discussions. RJ acknowledges the support of the SKA - South Africa and the National Research Foundation (NRF), as well as the hospitality of the Institute of Cosmology and Gravitation (ICG) at the University of Portsmouth where some of this work was carried out. DB and RN  are supported by the UK Science \& Technology Facilities Council (grant nos. ST/H002774/1 and ST/K0090X/1). The analysis was performed with the SCIAMA High Performance Computing cluster supported by the ICG, at the University of Portsmouth. We would like to thank the SCIAMA cluster administrator, Gary Burton, for all his help. Please contact the authors to request access to research materials discussed in this paper.

%\bibliography{../bibliographydb}

\end{document}